# Saturation magnetization and band gap tuning in BiFeO$_3$ nanoparticles via co–substitution of Gd and Mn

Mehedi Hasan [1,a)], M. A. Basith [2], M. A. Zubair [1], Md. Sarowar Hossain [3], Rubayyat Mahbub [1,4], M. A. Hakim [1], Md. Fakhrul Islam [1]

[1] *Department of Glass and Ceramic Engineering, Bangladesh University of Engineering and Technology, Dhaka 1000, Bangladesh*

[2] *Department of Physics, Bangladesh University of Engineering and Technology, Dhaka 1000, Bangladesh*

[3] *S. N. Bose National Centre for Basic Sciences, Salt Lake City, Kolkata, West Bengal 700098, India*

[4] *Department of Materials Science and Engineering, Carnegie Mellon university, 5000 Forbes avenue, Pittsburgh, PA 15213, USA*

[a)] Authors to whom correspondence should be addressed. Electronic mail: mhrizvi@gce.buet.ac.bd; rizvimme0711017@gmail.com

**Abstract:**

In this investigation, Gd and Mn co–doped Bi$_{0.85}$Gd$_{0.15}$Fe$_{1-x}$Mn$_x$O$_3$ (x=0.0–0.15) nanoparticles have been prepared to report the influence of co–substitution on their structural, optical, magnetic and electrical properties. Due to simultaneous substitution of Gd and Mn in BiFeO$_3$, the crystal structure has been modified from rhombohedral (*R3c*) to orthorhombic (*Pn2$_1$a*) and the Fe–O–Fe bond angle and Fe–O bond length have been changed. For Mn doping up to 10% in Bi$_{0.85}$Gd$_{0.15}$Fe$_{1-x}$Mn$_x$O$_3$ nanoparticles, the saturation magnetization (M$_s$) has been enhanced significantly, however, for a further increase of doping up to 15 %, the M$_s$ has started to reduce again. The co–substitution of Gd and Mn in BiFeO$_3$ nanoparticles also demonstrates a strong reduction in the optical band gap energy and electrical resistivity compared to that of undoped BiFeO$_3$.

1. Introduction:

Multiferroics are categorized as multifunctional materials which possess ferromagnetism/ antiferromagnetism and ferroelectricity simultaneously [1]. Multiferroic BiFeO$_3$ (BFO) is one of the rare single phase materials which exhibits multiferroicity above room temperature. Owing to its high ferroelectric Curie temperature ($T_c$ ~1103K) and antiferromagnetic Néel temperature ($T_N$ ~ 643K), BFO has attracted the attention of numerous researchers for its potential applications in spintronics devices [2]. In addition, BFO by virtue of its small band gap energy can utilize a wide spectrum of the sunlight in photo–induced applications pertaining to photovoltaic effect [3] and photocatalytic activity [4, 5].

In recent years, attention has been projected towards investigating the effects of simultaneous substitution of Bi and Fe in place of BFO to improve its multiferroic properties for multifunctional applications. Our previous investigations demonstrated that co–substitution of Bi and Fe sites of BFO by ions such as Gd and Ti can significantly improve structural, dielectric and magnetic properties of BFO ceramic in both bulk and nano scale [6, 7]. However, although the multiferroic properties of 10% Gd and Mn co–doped sub–micron BFO particles are studied a further investigation is claimed for different doping concentration in case of its nanostructured counterpart [8]. In addition, extensive investigations have also been carried out to tune the optical band gap of BFO ceramic by introducing this co–doping approach. Zhou et al. have reported that saturation magnetization and optical band gap increases in Sm and Mn co–doped BFO [9]. Similar increase in optical band gap is also reported for Y and Co co–doping of BFO [10]. Whereas, Kumar et al. have demonstrated that Ca and Ti co–doping reduces optical band gap of BFO [11]. Previous experiments have demonstrated that Gd substitution [12] increases band gap of BFO ceramic while Mn substitution [13] reduces. However, the effects of their (Gd-Mn) co-substitution on optical band gap energy of BFO nanoparticles have not been reported yet to the best of our knowledge. Therefore, in this investigation, we report the effects of co–substitution by Gd and Mn on structural, optical, magnetic and electrical properties of nanostructured BFO. A polar orthorhombic ($Pn2_1a$) modification of $Bi_{0.85}Gd_{0.15}Fe_{1-x}Mn_xO_3$ (x=0–0.15) nanoparticles with significantly enhanced magnetization and reduced band gap energy is observed.

2. Experimental:

We utilized sol–gel technique to synthesize $BiFeO_3$ (BFO), $Bi_{0.85}Gd_{0.15}FeO_3$ (BGFO), $Bi_{0.85}Gd_{0.15}Fe_{0.95}Mn_{0.05}O_3$ (BGFMO–5), $Bi_{0.85}Gd_{0.15}Fe_{0.9}Mn_{0.1}O_3$ (BGFMO–10) and $Bi_{0.85}Gd_{0.15}Fe_{0.85}Mn_{0.15}O_3$ (BGFMO–15) nanoparticles. Stoichiometric proportion of analytical grade pure Bi(NO$_3$)$_3$·5H$_2$O, Fe(NO$_3$)$_3$·9H$_2$O, along with Gd(NO$_3$)$_3$.6H$_2$O and Mn(NO$_3$)$_2$.4H$_2$O were dissolved in 400 ml deionized water. Citric acid and ethylene glycol were used as chelating and polymerization agents respectively. The detailed synthesis methodology to obtain BFO precursor xerogel has been described in our previous investigation [14]. The precursor xerogel powder and pelletized samples were annealed at 600°C for further characterization.

Powder X–ray diffraction (XRD, 3040–X'Pert PRO, Philips) of the synthesized nanoparticles was carried out within a 2θ range of 10° to 70° for the crystal structure analysis and phase identification. Particle size and morphology was observed adopting field emission scanning electron microscope (FESEM, JSM 7600, Jeol). Magnetic hysteresis measurements were carried out under applied field in the range of ± 15 kOe utilizing a vibrating sample magnetometer (VSM 7407, Lake Shore). The optical band gap energy of the synthesized samples was measured from diffused reflectance spectra using a UV–Vis spectrophotometer (UV–2600, Shimadzu). The dc resistivity and current density were investigated employing a ferroelectric loop tracer unit in conjunction with a 10kV external amplifier (Precision Multiferroic, Radiant Technologies, inc.).

3. **Results and discussion:**

Room temperature XRD pattern of the BFO, BGFO, BGFMO–5, BGFMO–10 and BGFMO–15 samples (Fig. 1) confirms the formation of crystalline BFO phase with some peaks (∗) associated with Bi$_{25}$FeO$_{40}$ impurity phase (ICDD 57594–58–8). The presence of small amount of other impurity phases cannot be ruled out during synthesis of these nanoparticles. Notably, the contribution of Bi$_{25}$FeO$_{40}$ impurity phase in magnetic ordering of BFO is negligible due to its paramagnetic nature at room temperature [15]. Full width at half maximum intensity (FWHM), β of the (102) diffraction peak was utilized in Scherrer equation d = kλ/β cos θ, to calculate the average crystallite size, *d* of BFO, BGFO, BGFMO–5, BGFMO–10 and BGFMO–15 nanoparticles, where k is the dimensionless constant with a typical value of ∼0.9, λ is the wavelength of Cu Kα radiation with the value of 1.5418 Å and θ is the Bragg angle of (102) diffraction peak. To investigate the changes in crystal structure and parameters, Rietveld refinement of the synthesized samples was carried out using FULLPROF Suite program. The refined structural parameters of the synthesized samples are enlisted in Table 1. The XRD pattern of BFO sample matches with single phase rhombohedral model (space group *R3c*). The coexistence of 80% rhombohedral (*R3c*) and 20% orthorhombic polar phase (*Pn2$_1$a*) is obtained for BGFO. Similarly, coexistence of 44% *R3c* and 56% *Pn2₁a* in case of BGFMO–5 and 28% *R3c* and 72% *Pn2₁a* in case of BGFMO–10 samples have been observed which are closely similar to Dy and Mn co–doped $Bi_{1-x}Dy_xFe_{1-x}Mn_xO_3$ multiferroics [16]. The *R3c* to *Pn2₁a* phase transition is found to increase with increasing Mn doping and single *Pn2₁a* phase is obtained for BGFMO–15. It is speculated that changes in chemical pressure and difference between ionic radii of dopants and substituted ions could play a role to modify the crystal structure of BFO. Moreover, it is observed that Fe–O–Fe bond angle and Fe–O bond length for rhombohedral, R3c phase (BFO) decreases from 153.904° and 1.8895Å to about 144.754° and 1.3157Å for orthorhombic, *Pn2₁a* phase (BGFMO–15). This change in structural parameters is expected to affect the magnetic and optical properties of BFO.

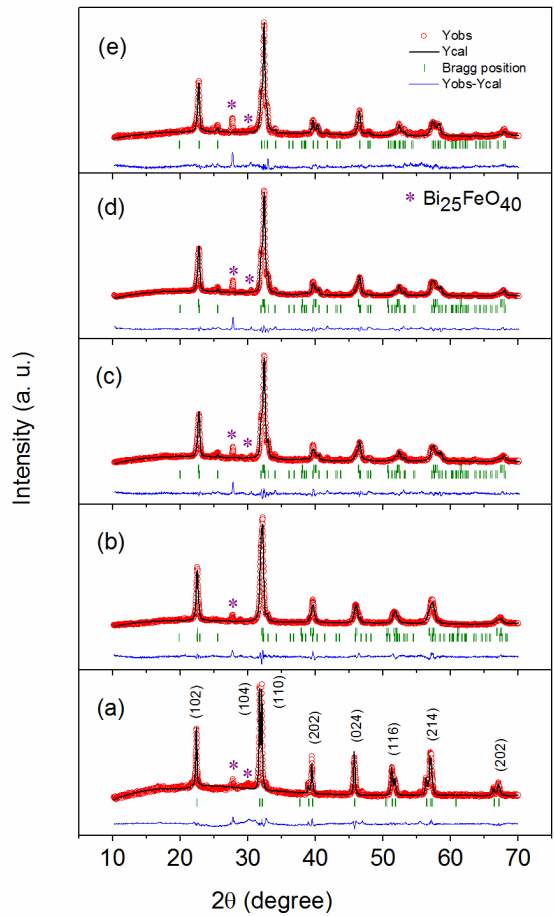

Fig. 1. The observed (Yobs), calculated (Ycal) XRD profiles and their difference (Yobs–Ycal) for (a) BFO, (b) BGFO, (c) BGFMO–5, (d) BGFMO–10 and (e) BGFMO–15 nanoparticles annealed at 600°C. The vertical marks below the profile show the Bragg position.

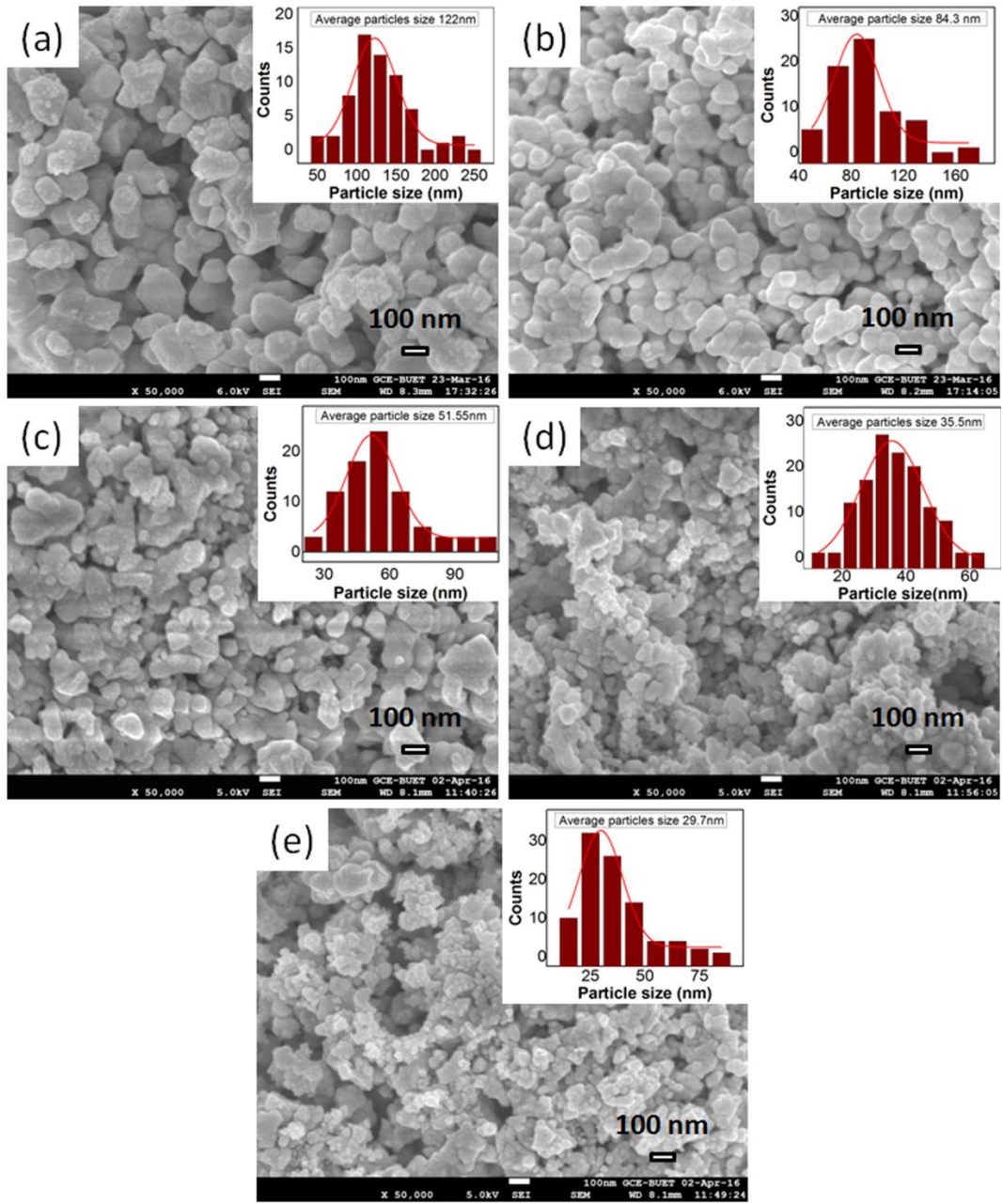

Fig. 2. FESEM micrographs of (a) BFO, (b) BGFO, (c) BGFMO–5, (d) BGFMO–10 and (e) BGFMO–15 nanoparticles. Inset: respective histograms regarding particle size distribution.

The FESEM micrographs and histograms regarding particle size distribution of BFO, BGFO, BGFMO–5, BGFMO–10 and BGFMO–15 are depicted in Fig.2 which demonstrate the effect of Gd and Mn co–doping on microstructure and particle size of BFO nanoparticles. The average crystallite sizes calculated from Scherrer formula are smaller than that of observed particle sizes from FESEM micrographs (Table 1) which suggests the presence of particle agglomeration effect [17, 18]. In sol-gel synthesis technique the complexing ligands of precursor xerogel cover nanocrystals and hinder particle agglomeration [19]. However, during annealing at 600°C the citrate complexing agents evaporate at above 300°C [14] and the ultra-fine BFO nanocrystals due to high surface energy form neck by solid state diffusion and evaporation-condensation process which leads to agglomeration and particle growth [20]. Nevertheless, both the crystallite size and particle size of Gd-Mn co-doped BFO nanoparticles decreases with increasing doping concentrations demonstrating the grain growth inhibition effect of Gd and Mn substitution. The differences in ionic radii between A-site $Bi^{3+}$ (1.03Å) and $Gd^{3+}$ (0.938 Å) ion and B-site $Fe^{3+}$ (0.645Å) and $Mn^{2+}$ (0.83), $Mn^{3+}$ (0.645 Å), $Mn^{4+}$ (0.53 Å) ions lead to a lattice distortion which hinder crystallite nucleation and thereby reduces particle size [21].

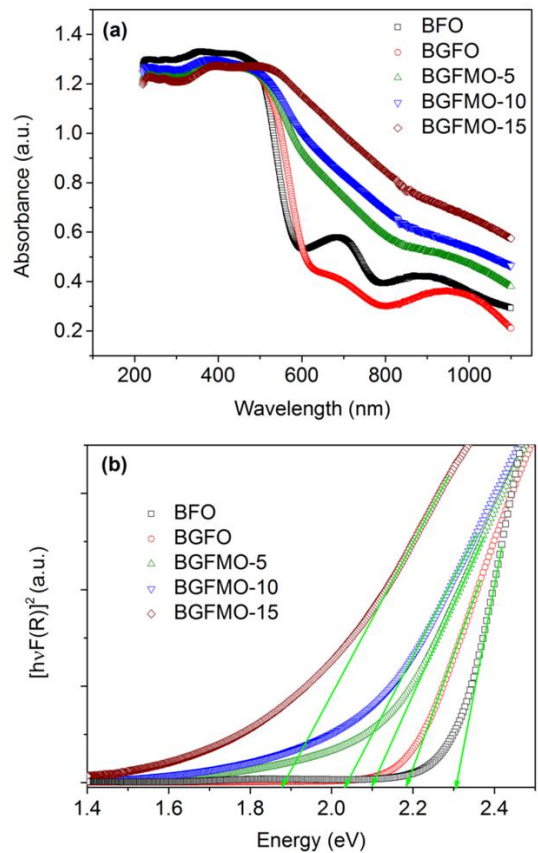

Fig. 3. (a) UV–Vis absorption spectra derived from diffused reflectance spectra of BFO, BGFO, BGFMO–5, BGFMO–10 and BGFMO–15. (b) $[hvF(R)]^2$ vs photon energy, hv plots to calculate band gap energy of the corresponding samples.

Fig. 3 (a) shows the absorbance vs. wavelength plot of undoped and doped BFO samples derived from diffused reflectance data using Kubelka–Munk [22] function which demonstrates chronological red shift by virtue of Gd and Mn doping. The diffused reflectance data was converted to Kubelka–Munk function given by $F(R)=\frac{(1-R)^2}{2R}$ (where R is the diffused reflectance value) to construct the $[F(R)hv]^2$ vs. $hv$ plots for the synthesized samples (Fig. 3 (b)). Here the intersection of the tangent line with $[F(R)hv]^2=0$ represents the optical band gap energy, $E_g$ [23] which corresponds to the energy difference between top of the valence band (O 2$p$) and bottom of the conduction band (Fe 3$d$) of BFO. The Fig. 3 (b) demonstrates that 15% Gd and Mn co–doping has appreciably reduced the optical band gap of BFO from 2.30 eV to about 1.87 eV. Notably, this reduced value is much smaller than the previously reported values for both undoped and doped BFO to the best of our knowledge [5, 9-12]. The band gap energies for undoped, Gd–doped and Gd–Mn co–doped nanoparticles are inserted in table 1. There may be several reasons rending behind this reduced band gap in the doped BFO samples. A first principles calculation has shown that due to reduced degree of hybridization associated with a stable electronic configuration (half occupation $4f^7 5d^0 6s^0$) of $Gd^{3+}$ ion, a unique energy level is formed in between Fe 3$d$ and O 2$p$ and therefore reduces effective band gap of Gd doped BFO [24]. Moreover, previous investigations have shown that the changes in Fe–O bond length and Fe–O–Fe bond angle by cation doping plays a critical role in modifying one–electron bandwidth (W) and hence band gap of BFO [25]. The empirical formula relating W with bond length and angle is, $W \propto \cos \omega / d_{Fe-O}^{3.5}$, where ω is ½ [π–(Fe–O–Fe)] and $d_{Fe-O}$ is the Fe–O bond length [26, 27]. The band gap is related with W as follows: $E_g=\Delta–W$, where Δ is the charge–transfer energy [26]. As the Fe–O bond length of orthorhombic phase is much smaller than rhombohedral phase (Table 1) [28], Mn–doping in $Bi_{0.85}Gd_{0.15}Fe_{1-x}Mn_xO_3$ nanoparticles may appreciably increase the value of W and hence decrease effective band gap energy with increasing orthorhombic phase. In addition, the 3$d$ conduction band edge of $Mn^{4+}$ ($E_{cb}$= –5.83 eV) is lower than $Fe^{3+}$ ($E_{cb}$= –4.78 eV) state [29] and brings holes in the $d$ band [30] which may reduce the effective energy gap between O 2$p$ valence band and Fe 3$d$ conduction band and hence decrease optical band gap width of $Bi_{0.85}Gd_{0.15}Fe_{1-x}Mn_xO_3$ nanoparticles.

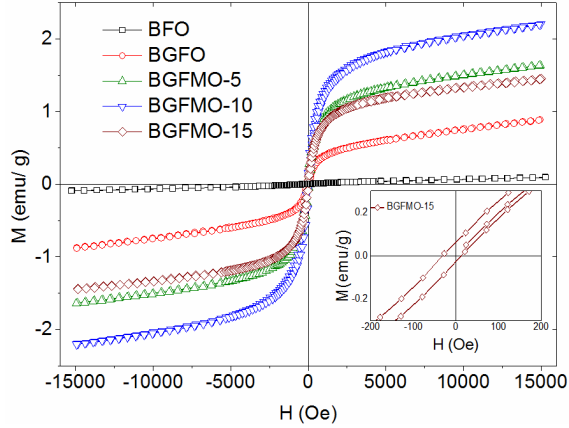

Fig. 4. Room temperature magnetic hysteresis loops of BFO, BGFO, BGFMO–5, BGFMO–10 and BGFMO–15 samples with inset showing magnified M–H loop of BGFMO–15.

Similar to the optical properties the magnetic properties of BFO nanoparticles are found to have greatly influenced by the structural transformation as a result of Gd and Mn co-doping. Fig. 4 depicts the magnetization vs. applied magnetic field (M–H) loops for BFO, BGFO, BGFMO–5, BGFMO–10 and BGFMO–15 nanoparticles at room temperature. The nearly linear magnetic field induced magnetization in case of BFO indicates its antiferromagnetic nature. For rhombohedral $R3c$ perovskite structure of BFO the Goodenough–Kanamori rules predict that the superexchange interaction between two $Fe^{3+}$ ions (high spin configuration $t_{2g}^3 e_g^2$) is antiferromagnetic and originate G–type spiral modulated spin stricture (SMSS) [31]. However, an enhancement in magnetization by Gd–doping at Bi–site and Mn–doping at Fe–site is evidenced in Fig. 4. The figure shows that 15% Gd–doping (BGFO) in BFO brings saturation magnetization ($M_s$) at 15 kOe to about 0.88 emu/g. Subsequently, 5% (BGFMO–5) and 10% (BGFMO–10) Mn–doping in BGFO has increased the $M_s$ to about 1.64 and 2.20 emu/g, respectively. The enhanced magnetization could be attributed to the following reasons. (i) The changes in bond angle and bond length (Table 1) by Gd and Mn doping in BFO may modify the tilting angle of $FeO_6$ octahedron and thereby suppress SMSS and trigger onset of magnetization [21, 32]. (ii) In the orthorhombic phases the spin canting originated by Dzyaloshinsky–Moriya interaction may introduce magnetization in the co–doped BFO [33, 34]. The growth of orthorhombic phase with increasing doping concentration may also contribute to the enhanced magnetization. (iii) Moreover, the substitution of non–magnetic $Bi^{3+}$ ( [Xe] $4f^{14}\ 5d^{10}\ 6s^2\ 6p^0$ ) ion by magnetic $Gd^{3+}$ ( [Xe] $4f^7$, theoretical magnetic moment 7 $\mu_B$) ion may be another reason behind this improved magnetization. In addition, Mn generally stay at multivalent (+2, +3 and +4) states in BFO [35, 36] and the substitution of $Fe^{3+}$ ([Ar] $3d^5$, 5 $\mu_B$) by $Mn^{2+}$ ( [Ar] $3d^5$, 5 $\mu_B$), $Mn^{3+}$ ( [Ar] $3d^4$, 4 $\mu_B$) and $Mn^{4+}$ ( [Ar] $3d^3$, 3 $\mu_B$) ion may cause incomplete compensation of antiferromagnetic SMSS and onset net magnetization. Furthermore, it is worth mentioning that Gd and Mn co–doping in BFO has significantly reduced particle size of the synthesized nanoparticles. The suppression of SMSS and contribution of uncompensated spins at the particle surface increases sharply with decreasing particle size of BFO nanoparticles [37, 38] and may subscribe to the enhanced magnetization with increasing doping concentration. The Ms values for undoped and co–doped BFO nanoparticles are enlisted in table 1.

The $M_s$ is found to increase gradually up to 10% Mn–doping in BGFMO–10, whereas for a further increase of doping concentration up to 15% in BGFMO –15 the $M_s$ reduces to about 1.44 emu/g. The complex interplay between cation exchange interactions would reduce magnetization of BGFMO–15. The antiferromagnetic nature of the Fe–O–Mn superexchange and [35] Gd–Mn exchange interaction [39] may compete with the ferromagnetic interaction associated with co–doping and hence reduce magnetization in BGFMO–15 nanoparticles. In addition, the M–H hysteresis loop of BGFMO–15 (inset of Fig. 4) demonstrates an asymmetric shift towards the field axis which is a signature of exchange bias effect and also indicates the presence of competing antiferromagnetic and ferromagnetic interaction [40].

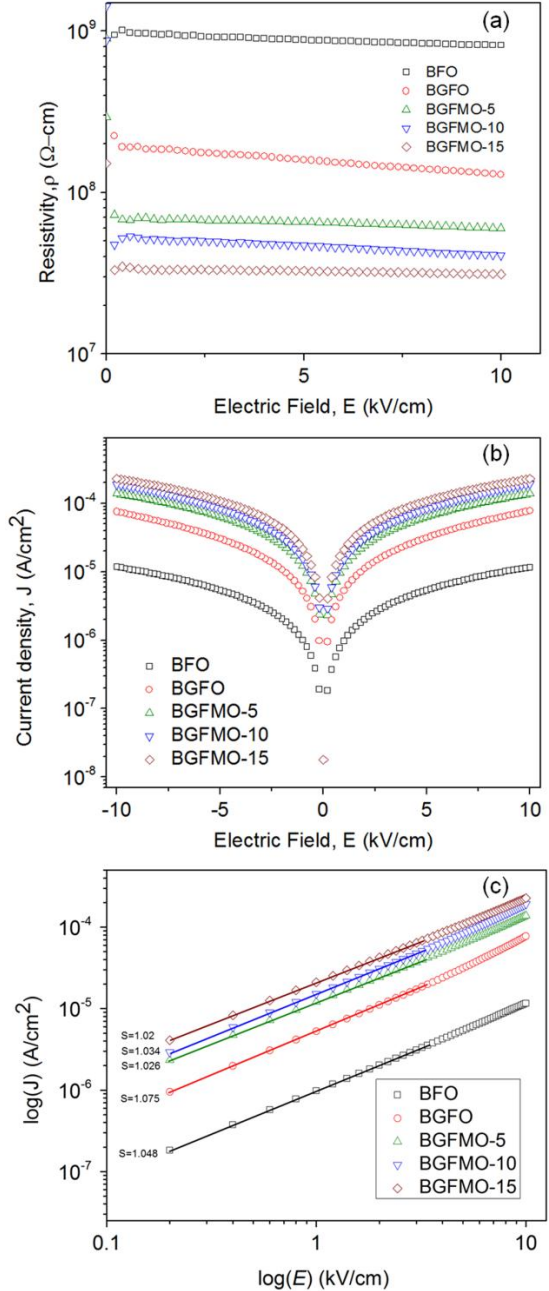

Fig. 5. (a) Resistivity, $\rho$ (Ω–cm) vs. Electric Field, $E$ (kV/cm) plots (b) Current density, $J$ (A/cm$^2$) vs. Electric Field, $E$ (kV/cm) plots and (c) log($J$) vs Log($E$) plots of BFO, BGFO, BGFMO–5, BGFMO–10 and BGFMO–15 ceramics.

The dc resistivity, $\rho$ vs electric field, $E$ and leakage current density, $J$ vs electric filed, $E$ plots of BFO, BGFO, BGFMO–5, BGFMO–10 and BGFMO–15 pelletized sample are shown in Fig 5.(a) and (b), respectively. To investigate the conduction mechanism and origin of leakage current of the co–doped samples log(*J*) vs log(*E*) plots have been constructed and are depicted in Fig. 5(c). The linearity of the log(J) vs log(E) plots with slope ~1 indicates that ohmic conduction mechanism prevails up to 10 kV/cm for the synthesized samples, whereas nonlinear space charge limiting conduction mechanism appears at higher applied field [41, 42]. In the ohmic conduction mechanism the leakage current density is associated with the thermally generated free electrons and can be expressed as $J=e\mu N_e E$, where $e$, $\mu$, $N_e$, and $E$ are the charge of a electron, carrier mobility of free electron, density of the thermally stimulated electrons, and applied electric field, respectively [41, 42]. The concentration of free electron–hole pair, $N_e$ increases sharply with reducing optical band gap energy and thereby raise leakage current density [43]. Therefore, the observed increased conductivity of synthesized samples with increasing Gd and Mn co–doping concentration may be attributed to the reduced energy gap between Fermi level and conduction band. Hence the reduced effective band gap of the co–doped samples as discussed earlier corroborates the observed increased conductivity and reduced resistivity of the Gd-Mn co-doped samples as demonstrated in Fig. 5. Moreover, the electrical conductivity of nanocrystalline oxide materials shows strong dependence on particle size and increases with reducing particle size. It is supposed that the donor effect of grain boundary and easy path for ionic diffusion promote electrical conductivity of nanocrystalline BFO with reducing particle size [38, 44]. The decreased particle size by means of Gd-Mn co-doping which is an important factor behind enhanced magnetization however, may contribute to the increased electrical conductivity of nanocrystalline BFO samples with increasing doping concentration.

4. **Conclusions:**

Structural, optical, magnetic and electrical properties of sol–gel derived Gd and Mn co–doped BFO were investigated. Co-substitution with Gd and Mn leads to structural transformation from rhombohedral (*R3c*) to orthorhombic (*Pn2$_1$a*) symmetry which was found to have great effect on optical band gap energy, magnetization and electrical conductivity of BFO. The significantly enhanced ferromagnetic ordering was achieved in the co–doped samples while antiferromagnetic nature is the primary weakness of undoped BFO regarding its potential multifunctional applications. The ohmic conduction mechanism was found to control electrical conductivity of co-doped BFO samples upto 10kV/cm which implies little contribution from space charge at low applied electric filed. The reduced optical band gap and electrical resistivity by virtue of Gd and Mn co-doping may find potential applications in the optoelectronic devices.

**Acknowledgements:**

The authors would like to thank Ministry of education, the Peoples Republic of Bangladesh for its financial support. The authors also would like to acknowledge Infrastructure Development Company Limited (IDCOL), Bangladesh for providing optical measurement facility under a research grant.

**References**

[1] W. Eerenstein, N.D. Mathur, J.F. Scott ' *Multiferroic and magnetoelectric materials.* ' Nature, 442 (2006) 759-765.

[2] G. Catalan, J.F. Scott ' *Physics and Applications of Bismuth Ferrite.* ' Advanced Materials, 21 (2009) 2463–2485.

[3] S.Y. Yang, L.W. Martin, S.J. Byrnes, T.E. Conry, S.R. Basu, D. Paran, L. Reichertz, J. Ihlefeld, C. Adamo, A. Melville, Y.-H. Chu, C.-H. Yang, J.L. Musfeldt, D.G. Schlom, J.W. Ager, R. Ramesh ' *Photovoltaic effects in BiFeO$_3$.* ' Applied Physics Letters, 95 (2009) 062909.

[4] F. Gao, X.Y. Chen, K.B. Yin, S. Dong, Z.F. Ren, F. Yuan, T. Yu, Z.G. Zou, J.M. Liu ' *Visible-Light Photocatalytic Properties of Weak Magnetic BiFeO3 Nanoparticles.* ' Advanced Materials, 19 (2007) 2889-2892.

[5] X. Bai, J. Wei, B. Tian, Y. Liu, T. Reiss, N. Guiblin, P. Gemeiner, B. Dkhil, I. C. Infante ' *Size Effect on Optical and Photocatalytic Properties in BiFeO3 Nanoparticles.* ' The Journal of Physical Chemistry C, 120 (2016) 3595-3601.

[6] M.A. Basith, O. Kurni, M.S. Alam, B.L. Sinha, B. Ahmmad ' *Room temperature dielectric and magnetic properties of Gd and Ti co-doped BiFeO$_3$ ceramics.* ' Journal of Applied Physics, 115 (2014) 024102.

[7] M.A. Basith, D.-T. Ngo, A. Quader, M.A. Rahman, B.L. Sinha, B. Ahmmad, F. Hirosed, K. Mølhaveb ' *Simple top-down preparation of magnetic Bi$_{0.9}$Gd$_{0.1}$Fe$_{1−x}$Ti$_x$O$_3$ nanoparticles by ultrasonication of multiferroic bulk material.* ' Nanoscale, 6 (2014) 14336-14342.

[8] P. Tang, D. Kuang, S. Yang, Y. Zhang '*Structural, morphological and multiferroic properties of the hydrothermally grown gadolinium (Gd) and manganese (Mn) doped sub-micron bismuth ferrites.*' Journal of Alloys and Compounds, 656 (2016) 912-919.

[9] W. Zhou, H. Deng, H. Cao, J. He, J. Liu, P. Yang, J. Chu '*Effects of Sm and Mn co-doping on structural, optical and magnetic properties of BiFeO3 films prepared by a sol–gel technique.*' Materials Letters, 144 (2015) 93-96.

[10] D. Kuang, P. Tang, X. Wu, S. Yang, X. Ding, Y. Zhang '*Structural, optical and magnetic studies of (Y, Co) co-substituted BiFeO3 thin films.*' Journal of Alloys and Compounds, 671 (2016) 192-199.

[11] P. Kumar, M. Kar '*Effect of structural transition on magnetic and optical properties of Ca and Ti co-substituted BiFeO$_3$ ceramics.*' Journal of Alloys and Compounds, 584 (2014) 566-572.

[12] R. Guo, L. Fang, W. Dong, F. Zheng, M. Shen '*Enhanced Photocatalytic Activity and Ferromagnetism in Gd Doped BiFeO3 Nanoparticles.*' The Journal of Physical Chemistry C, 114 (2010) 21390-21396.

[13] S. Chauhan, M. Kumar, S. Chhoker, S.C. Katyal, H. Singh, M. Jewariya, K.L. Yadav '*Multiferroic, magnetoelectric and optical properties of Mn doped BiFeO3 nanoparticles.*' Solid State Communications, 152 (2012) 525–529.

[14] M. Hasan, M.F. Islam, R. Mahbub, M.S. Hossain, M.A. Hakim '*A soft chemical route to the synthesis of BiFeO$_3$ nanoparticles with enhanced magnetization.*' Materials Research Bulletin, 73 (2016) 179-186.

[15] Y. Chena, Q. Wua, J. Zhaoa '*Selective synthesis on structures and morphologies of BixFeyOz nanomaterials with disparate magnetism through time control.*' Journal of Alloys and Compounds, 13 (2009) 599–604.

[16] S.N. Tripathy, D.K. Pradhan, K.K. Mishra, S. Sen, R. Palai, M. Paulch, J.F. Scott, R.S. Katiyar, D.K. Pradhan '*Phase transition and enhanced magneto-dielectric response in BiFeO$_3$-DyMnO$_3$ multiferroics.*' Journal of Applied Physics, 117 (2015) 144103.

[17] J. Liu, L. Fang, F. Zheng, S. Ju, M. Shen '*Enhancement of magnetization in Eu doped BiFeO3 nanoparticles.*' Applied Physics Letters, 95 (2009) 022511.

[18] S.M. Selbach, T. Tybell, M.-A. Einarsrud, T. Grande '*Size-Dependent Properties of Multiferroic BiFeO3 Nanoparticles.*' Chemistry of Materials, 19 (2007) 6478-6484.

[19] M.M. Bagheri-Mohagheghi, N. Shahtahmasebi, M.R. Alinejad, A. Youssefi, M. Shokooh-Saremi ' *The effect of the post-annealing temperature on the nano-structure and energy band gap of SnO2 semiconducting oxide nano-particles synthesized by polymerizing–complexing sol–gel method.*' Physica B: Condensed Matter, 403 (2008) 2431-2437.

[20] B.A. Hernandez, K.-S. Chang, E.R. Fisher, P.K. Dorhout ' *Sol−Gel Template Synthesis and Characterization of BaTiO3 and PbTiO3 Nanotubes.*' Chemistry of Materials, 14 (2002) 480-482.

[21] P. Kumar, N. Shankhwar, A. Srinivasan, M. Kar ' *Oxygen octahedra distortion induced structural and magnetic phase transitions in $Bi_{1−x}Ca_xFe_{1−x}Mn_xO_3$ ceramics.*' Journal of Applied Physics, 117 (2015) 194103.

[22] P. Kubelka, F. Munk ' *Ein Beitrag Zur Optik Der Farbanstriche.* ' Zeitschrift für Technische Physik, 12 (1931) 593-601.

[23] P.S.V. Mocherla, C. Karthik, R. Ubic, M.S.R. Rao, C. Sudakar ' *Tunable bandgap in $BiFeO_3$ nanoparticles: The role of microstrain and oxygen defects.*' Applied Physics Letters, 103 (2013) 022910.

[24] N. Gao, W. Chen, R. Zhang, J. Zhang, Z. Wu, W. Mao, J. Yang, X.a. Li, W. Huang '*First principles investigation on the electronic, magnetic and optical properties of $Bi_{0.8}M_{0.2}Fe_{0.9}Co_{0.1}O_3$ (M = La, Gd, Er, Lu).*' Computational and Theoretical Chemistry, 1084 (2016) 36-42.

[25] Z. Zhang, P. Wu, L. Chen, J. Wang ' *Systematic variations in structural and electronic properties of $BiFeO_3$ by A-site substitution.*' Applied Physics Letters, 96 (2010) 012905.

[26] M. Medarde, J. Mesot, P. Lacorre, S. Rosenkranz, P. Fischer, K. Gobrecht ' *High-pressure neutron-diffraction study of the metallization process in $PrNiO_3$.*' physical Review B, 52 (1995) 9248-9258.

[27] P.G. Radaelli, G. Iannone, M. Marezio, H.Y. Hwang, S.-W. Cheong, J.D. Jorgensen, D.N. Argyriou ' *Structural effects on the magnetic and transport properties of perovskite $A_{1−x}A'_xMnO_3$ (x=0.25, 0.30).*' physical Review B, 56 (1997) 8265-8276.

[28] H.M. Tütüncü, G.P. Srivastava ' *Electronic structure and lattice dynamical properties of different tetragonal phases of $BiFeO_3$.*' Physical Review B, 78 (2008) 235209.

[29] Y. Xu, M.A.A. Schoonen ' *The absolute energy positions of conduction and valence bands of selected semiconducting minerals.*' American Mineralogist, 85 (2000) 543–556.

[30] J. Rodríguez-Carvajal, M. Hennion, F. Moussa, A.H. Moudden, L. Pinsard, A. Revcolevschi ' *Neutron-diffraction study of the Jahn-Teller transition in stoichiometric LaMnO$_3$.*' Physical Review B, 57 (1998) R3189-R3192.

[31] I. Sosnowska, T.P. Neumaier, E. Steichele ' *Spiral magnetic ordering in bismuth ferrite.*' Journal of Physics C: Solid State Physics, 15 (1982) 4835.

[32] P. Kumar, M. Kar ' *Tuning of net magnetic moment in BiFeO3 multiferroics by co-substitution of Nd and Mn.*' Physica B: Condensed Matter, 448 (2014) 90-95.

[33] V.A. Khomchenko, L.C.J. Pereira, J.A. Paixão '*Substitution-driven structural and magnetic phase transitions in Bi$_{0.86}$(La, Sm)$_{0.14}$FeO$_3$ system.*' journal of Physics D: Applied Physics, 44 (2011) 185406.

[34] I.A. Sergienko, E. Dagotto ' *Role of the Dzyaloshinskii-Moriya interaction in multiferroic perovskites.*' Physical Review B, 73 (2006) 094434.

[35] Q. Xu, Y. Sheng, M. Khalid, Y. Cao, Y. Wang, X. Qiu, W. Zhang, M. He, S. Wang, S. Zhou, Q. Li, D. Wu, Y. Zhai, W. Liu, P. Wang, Y.B. Xu, J. Du ' *Magnetic interactions in BiFe$_{0.5}$Mn$_{0.5}$O$_3$ films and BiFeO$_3$/BiMnO$_3$ superlattices.*' Scientific Reports, 5 (2015) 09093.

[36] Y. Yoneda, Y. Kitanaka, Y. Noguchi, M. Miyayama ' *Electronic and local structures of Mn-doped BiFeO$_3$ crystals.* Physical Review B, 2012.' Physical Review B, 86 (2012) 184112.

[37] T.-J. Park, C.G. Papaefthymiou, J.A. Viescas, R.A. Moodenbaugh, S.S. Wong ' *Size-dependent magnetic properties of single-crystalline multiferroic BiFeO3 nanoparticles.*' Nano Letters, 7 (2007) 766-772.

[38] M. Hasan, M.A. Hakim, M.A. Basith, M.S. Hossain, B. Ahmmad, M.A. Zubair, A. Hussain, M.F. Islam ' *Size dependent magnetic and electrical properties of Ba-doped nanocrystalline BiFeO3.* AIP Advances. ' AIP Advances, 6 (2016) 035314.

[39] J. Hemberger, S. Lobina, H.-A.K.v. Nidda, N. Tristan, V.Y. Ivanov, A.A. Mukhin, A.M. Balbashov, A. Loidl ' *Complex interplay of 3d and 4f magnetism in La$_{1-x}$Gd$_x$MnO$_3$.*' Physical Review B, 70 (2004) 024414.

[40] M.A. Basith, F.A. Khan, B. Ahmmad, S. Kubota, F. Hirose, D.-T. Ngo, Q.-H. Tran, K. Mølhave ' *Tunable exchange bias effect in magnetic Bi$_{0.9}$Gd$_{0.1}$Fe$_{0.9}$Ti$_{0.1}$O$_3$ nanoparticles at temperatures up to 250 K .*' journal of Applied Physics, 118 (2015) 023901.

[41] C. Wang, M. Takahashi, H. Fujino, X. Zhao, E. Kume, T. Horiuchi, S. Sakai ' *Leakage current of multiferroic (Bi0.6Tb0.3La0.1)FeO3 thin films grown at various oxygen pressures by pulsed laser deposition and annealing effect.*' Journal of Applied Physics, 99 (2006) 054104.

[42] C.M. Raghavan, J.W. Kim, T.K. Song, S.S. Kim ' *Microstructural, electrical and ferroelectric properties of BiFe0.95Mn0.05O3 thin film grown on Ge-doped ZnO electrode.*' Materials Research Bulletin, 74 (2016) 164-168.

[43] M. Gratzel ' *Photoelectrochemical cells.*' Nature, 414 (2001) 338-344.

[44] A. Tschöpe, E. Sommer, R. Birringer ' *Grain size-dependent electrical conductivity of polycrystalline cerium oxide: I. Experiments.*' Solid State Ionics, 139 (2001) 255-265.

Tables

Table 1: Refined structural parameters, calculated crystallite sizes (d), calculated average particle sizes (D) from FESEM micrographs along with saturation magnetization, $M_s$ at 15 kOe and optical band gap energy, $E_g$ values of BFO, BGFO, BGFMO–5, BGFMO–10 and BGFMO–15 nanoparticles

| Sample | BFO | BGFO | | BGFMO–5 | | BGFMO–10 | | BGFMO–15 |
|---|---|---|---|---|---|---|---|---|
| d (nm) | 86 | 44 | | 32 | | 27 | | 24 |
| D (nm) | 122 | 84 | | 52 | | 36 | | 30 |
| a (Å) | 5.5747 | 5.5542 | 5.5717 | 5.5502 | 5.5975 | 5.5544 | 5.5976 | 5.5872 |
| b (Å) | 5.5747 | 5.5542 | 7.9095 | 5.5502 | 7.8069 | 5.5544 | 7.8038 | 7.7987 |
| c (Å) | 13.8616 | 13.7692 | 5.4412 | 13.5965 | 5.4278 | 13.5055 | 5.4409 | 5.4518 |
| Volume (Å³) | 373.2020 | 367.8594 | 239.7910 | 362.7169 | 237.1925 | 360.8450 | 238.51 | 237.5497 |
| Fe–O–Fe bond angle(°) | 153.904 | 149.3108 | 144.7539 | 152.825 | 143.1825 | 148.0951 | 143.9682 | 144.75391 |
| Fe–O bond length (Å) | 1.8895 | 1.8261 | 1.4943 | 1.6825 | 1.1760 | 1.8756 | 1.2639 | 1.3157 |
| Bi–O bond length (Å) | 2.2120 | 2.2071 | 2.1353 | 2.2417 | 2.1655 | 2.2262 | 2.2284 | 2.2684 |
| $\chi^2$ | 5.67 | 3.27 | | 3.92 | | 2.98 | | 3.56 |
| $M_s$ (emu/g) | 0.09 | 0.88 | | 1.64 | | 2.20 | | 1.44 |
| $E_g$ (eV) | 2.30 | 2.18 | | 2.10 | | 2.03 | | 1.87 |